\documentclass[a4paper]{jpconf}

\usepackage{graphicx}
\usepackage{lmodern}
\usepackage{lmodern}
\usepackage[T1]{fontenc}
\usepackage[latin9]{inputenc}
\usepackage{color}
\usepackage{amsmath}
\usepackage{babel}
\usepackage{revsymb}
\usepackage{cite}

\begin{document}

\global\long\def\l{\mathcal{L}}%
\global\long\def\d{\mathrm{d}}%

\title{A model of non-minimally coupled gravitation and electromagnetism
in (1+2) dimensions}

\author{K\i van\c{c} \.{I}. \"Unl\"ut\"urk$^1$ and Cem Yeti\c{s}mi\c{s}o\u{g}lu$^2$}

\address{Department of Physics, Ko\c{c} University, 34450, Sar\i yer, \.{I}stanbul, Turkey}

\ead{$^1$kunluturk17@ku.edu.tr \textrm{and} $^2$cyetismisoglu@ku.edu.tr}

\begin{abstract}
Following earlier works of Dereli and collaborators, we study a three dimensional toy model where we extend the topologically massive gravity with electrodynamics by the most general $RF^2$-type non-minimal coupling terms. Here $R$ denotes the possible curvature terms and $F$ denotes the electromagnetic 2-form. We derive the variational field equations and look for exact solutions on constant negative curvature space-times with a constant, self-dual electromagnetic field. The notion of self-dual electromagnetic fields in three dimensions is introduced by Dereli and collaborators in the study of exact solutions of models with gravity-electromagnetism couplings. We note the conditions that the parameters of the model have to satisfy for these self-dual solutions to exist.
\end{abstract}

\section{Introduction}
Three dimensions has the virtue of providing many interesting toy models. One of the main reasons these toy models are investigated is that these toy models help us gain insight to their more complicated four dimensional analogues. For instance, there is a plethora of three dimensional gravitational models which are studied to better understand the quantum gravity problem in four dimensions, see for instance \cite{deser1982topologically1, deser1988topologically2, Witten1, Witten2, HorneWitten, LiStrominger, Bergshoeff1, Bergshoeff2, Tekin,  carlip1998}. Although three dimensional general relativity with cosmological constant has no local degrees of freedom \cite{Deser1983constant, Deser1983flat}, one can obtain intriguing models that possess propagating degrees of freedom by adding a Lorentz Chern-Simons term to the action, one of the earliest examples being the Topologically Massive Gravity (TMG) theory \cite{deser1982topologically1, deser1988topologically2}. On the one hand its properties have been studied extensively: it admits a celebrated black hole solution \cite{banados1992} and a unitary quantum theory of gravity for the wrong sign of the Einstein-Hilbert term \cite{brown1986central} which is also renormalizable \cite{Oda}. On the other hand, different generalizations that contain additional fields have also been studied, e.g. in \cite{Dengiz:2012jb, DeserKay, DereliSarioglu1, DereliSarioglu2}.\\

\noindent In this manuscript we are going to study a generalisation of topologically massive gravity with electrodynamics that contains $RF^2$-type non-minimal couplings between electrodynamics and gravity. Similar toy models in four dimensions has been considered in \cite{DereliSert1, DereliSert2, DereliSert3, DereliSert4, DereliSenikoglu} where in general a single $RF^2$-type interaction is added to the Einstein-Maxwell theory. Although four dimensional Einstein-Maxwell theory, where the electromagnetic field is minimally coupled to gravity, is classically a well-established theory, non-minimal coupling terms between curvature and electrodynamics may be introduced, motivated by perturbative QED on curved backgrounds, in particular coming from vacuum polarization. On the tree level, photons propagate along null geodesics, whereas at the 1-loop level, due to vacuum polarization, a photon may exist as a virtual electron-positron pair. Consequently, photons attain a size at the order of magnitude of the Compton wavelength of the electron. Therefore, photon propagation can be influenced by the space-time curvature where a curved space-time acts as an optically active medium. Such effects can be conveniently described by non-minimal couplings of the electromagnetic field to gravity \cite{drummond1980}. Indeed, the effective
action density that arises in QED at the 1-loop level is given by
\cite{drummond1980} 
\begin{align*}
\frac{\gamma_{1}}{2} & R_{ab}\wedge F^{ab} \star F +\frac{\gamma_{2}}{2}Ric_{a}\wedge \iota^{a}F\wedge\star F +\frac{\gamma_{3}}{2}RF\wedge\star F
+\frac{\gamma_{4}}{2}\d\star F\wedge\star\left(\d\star F\right),
\end{align*}
where the coupling constants $\{\gamma_i\}$ depend on the fine structure constant and mass of the electron. The notation used here will be explained in detail shortly. \\

\noindent Motivated by these interactions, in \cite{DereliSert1, DereliSert2, DereliSert3, DereliSert4, DereliSenikoglu} the authors consider various extensions of Einstein-Maxwell theory with $RF^{2}$-type interaction terms and look for solutions. Particularly authors, consider an $R_{ab} \wedge F^{ab}\star F$ coupling term and show that a class of exact non-trivial pp-wave solutions exist in \cite{DereliSert1}; add coupling terms of the form $Y(R)F\wedge \star F$ and show that static, spherically symmetric solutions exist; look at exact solutions of anisotropic inflation for the same coupling in \cite{DereliSert4}. Finally in \cite{DereliSenikoglu} authors consider $RF \wedge\star F$ and $RF \wedge F$ couplings and show that there exists a charge screening solution (with vanishing total charge) for this model. That is, a spherically symmetric, asymptotically de Sitter solution with a constant electromagnetic field 2-form exists, provided that the coupling constant for the non-minimal coupling term is related to the radius of space-time in a certain way. Here, we extend three dimensional topologically massive gravity with electrodynamics to a model that contains all of the non-minimal coupling terms coming from QED. It is noteworthy to see similar effects coming from the non-minimal coupling terms of this toy model. We derive the field equations using first order constrained variational principle. We then study an exact class of AdS solutions with a constant self-dual electromagnetic field. The notion of a self-dual electromagnetic 2-form in three dimensions is introduced for the study of exact solutions of Einstein-Maxwell-Chern-Simons \cite{DereliObukhov} and TMG with electrodynamics \cite{DereliSarioglu1, DereliSarioglu2}. Here, what we mean by self-duality\footnote{In three dimensions, electromagnetic 2-form cannot be self-dual in the sense that it cannot be in an eigenspace of the Hodge duality operator.} is that the Faraday 2-form contains a single electric field component and a magnetic field whose magnitudes are equal. We show that similar to the result in \cite{DereliSenikoglu}, our model admits exact constant self-dual solutions on AdS space-time given that the coupling constants of the model are constrained by two algebraic relations. \\

\noindent The organization of the paper is as follows. We start by giving our notations and conventions. Then, in the second section, we give the action formulation of our model and derive the variational field equations. In the third section we look for the exact solutions to the variational field equations and in the final section we conclude with some final remarks.  \\

\noindent \textbf{Notations and Conventions:} We will be using the language of differential forms on three dimensional Riemann-Cartan manifolds. The metric of space-time is expressed as $g=\eta_{ab}e^a \otimes e^b$ where $\eta_{ab}=\text{diag}(-,+,+)$ and $\{e^a\}$ denotes the set-of $g$-orthonormal co-frames which are dual to the $g$-orthonormal frames $\{X_a\}$. The frame indices are raised and lowered by the use of $\eta_{ab}$. The operations $\star:\Omega^pM\to\Omega^{3-p}M$, $\d: \Omega^pM\to\Omega^{p+1}M$ and $\iota_a:\Omega^pM\to\Omega^{p-1}M$ denote the Hodge duality with respect to the metric $g$, exterior differentiation and interior product with respect to a frame $X_a$ operations, respectively. We use the shorthand notation $e^{ab\cdots}=e^a\wedge e^b\wedge \cdots$ and $\iota_{ab\cdots} = \iota_a \iota_b \cdots$. \\

\noindent A linear metric connection is specified by a set of 1-forms $\{\omega^a_{\ b}\}$ taking values in the three dimensional Lorentz algebra. The torsion and curvature 2-forms are defined via the first and second structure equations as
\begin{align}
T^a&= \d e^a +   \omega^a_{\ b} \wedge e^b, \\
R^a_{\ b}&=\d \omega^a_{\ b} + \omega^a_{\ c} \wedge \omega^c_{\ b},
\end{align}
respectively. They satisfy the Bianchi identities 
\begin{align}
D T^a &=   R^a_{\ b} \wedge e^b, \\
DR^a_{\ b}&=0,
\end{align}
where $D:\Omega^pM\to\Omega^{p+1}M$ denotes the exterior covariant derivative operator with respect to the linear connection. The Ricci 1-forms $\{Ric_a\}$ and the curvature scalar $R$ are obtained as contractions of the curvature 2-forms, that is
\begin{equation}
Ric_a = \iota_b R^b_{\ a} ,\qquad R= \iota^a Ric_a.
\end{equation}
In three dimensions, the curvature 2-forms, Ricci 1-forms and the curvature scalar are related to each other by the identity
\begin{equation}
R_{ab} = \epsilon^c_{\ ab} \star Ric_c +\frac{R}{2} e_a \wedge e_b, \label{curvature identity}
\end{equation}
where $\epsilon_{abc}$ is the totally anti-symmetric Levi-Civita symbol with $\epsilon_{012}=1$.

\section{Action and variational field equations}

The action functional for our model 
\begin{equation}
S[A,e^{a},\omega_{\ b}^{a},\lambda^{a}]=\int_{M}\mathcal{L}
\end{equation}
depends on four independent variables: the electromagnetic potential
1-form $A$ with Faraday 2-form $F=\d A$, co-frame 1-forms $\{e^{a}\}$,
connection 1-forms $\{\omega_{\ b}^{a}\}$ and Lagrange multiplier
1-forms $\{\lambda^{a}\}$. The Lagrangian 3-form can be decomposed
into four parts as
\begin{equation}
\l=\l_{\text{grav}}+\l_{\text{EM}}+\l_{\text{coup}}+\l_{\text{constraint}},\label{eq: lagrangian of the theory}
\end{equation}
where $\l_{\text{grav}}$ and $\l_{\text{EM}}$ denote the action
densities involving solely gravitational and electrodynamical terms,
$\l_{\text{coup}}$ denotes the non-minimal coupling terms and $\l_{\text{constraint}}$
is the constraint term which will set the torsion 2-forms to zero. \\

\noindent The gravitational action density 
\begin{equation}
\l_{\text{grav}}=\frac{1}{\mu}\left(\omega_{\phantom{a}b}^{a}\wedge\d\omega_{\phantom{a}a}^{b}+\frac{2}{3}\omega_{\phantom{a}b}^{a}\wedge\omega_{\phantom{a}c}^{b}\wedge\omega_{\phantom{a}a}^{c}\right)+\frac{1}{2}R_{ab}\wedge\star e^{ab}-\Lambda\star1
\end{equation}
contains the usual Einstein-Hilbert term with the cosmological constant
$\Lambda$. Furthermore, we also couple the Lorentz Chern-Simons
term with the coupling constant $\mu$. The electromagnetic action
density 
\begin{equation}
\l_{\text{EM}}=-\frac{1}{2}F\wedge\star F-\frac{m}{2}A\wedge F,
\end{equation}
contains the Maxwell density and Maxwell Chern-Simons term with coupling
constant $m$. For the couplings between gravitational and electromagnetic
fields, we consider the non-minimal coupling terms of type $RF^{2}$,
i.e.
\begin{align}
\l_{\text{coup}}&=\frac{\gamma_{1}}{2}F^{ab}R_{ab}\wedge\star F+\frac{\gamma_{2}}{2}Ric_{a}\wedge \iota^{a}F\wedge\star F+\frac{\gamma_{3}}{2}RF\wedge\star F \nonumber\\
&=\left( \frac{\gamma_{1}}{2}+\frac{\gamma_{2}}{4} \right)F^{ab}R_{ab}\wedge\star F +\left( \frac{\gamma_{2}}{4}+\frac{\gamma_{3}}{2} \right)RF\wedge\star F,
\end{align}
where in the last equality we make use of the curvature identity (\ref{curvature identity}) to get rid of the term proportional to Ricci 1-forms for technical ease. Finally, the constraint term reads 
\begin{equation}
\l_{\text{constraint}}=\lambda_{a}\wedge T^{a}.
\end{equation}

\noindent Upon the variation of the Lagrange multiplier 1-forms $\{\lambda_{a}\}$,
we obtain that torsion 2-forms vanish identically, that is $T^{a}=0$.
This ensures that our connection 1-forms are the unique Levi-Civita
connections 1-forms. Therefore in what follows, it is understood that
all curvatures and covariant derivatives are to be calculated using
the Levi-Civita connection 1-forms. The variation with respect to
the electromagnetic potential 1-form yield modified Maxwell's
equations 
\begin{align}
0 & =-\d\star F-mF+\left(\frac{\gamma_{1}}{2}+\frac{\gamma_{2}}{4}\right)\d\Big[R\star F-2\left(\iota^{a}\star F\right)Ric_{a}+F_{ab}\star R^{ab}\Big]\nonumber \\
 & \quad\quad+\left(\frac{\gamma_{2}}{4}+\frac{\gamma_{3}}{2}\right)\d\left(2R\star F\right).\label{eq: modified maxwell}
\end{align}
The variation with respect to the connection 1-forms $\{\omega^{ab}\}$
yield 
\begin{align}
\frac{1}{2}\left(e_{a}\wedge\lambda_{b}-e_{b}\wedge\lambda_{a}\right) & =-\frac{2}{\mu}R_{ab}+\frac{1}{2}D\star e_{ab}+\left(\frac{\gamma_{1}}{2}+\frac{\gamma_{2}}{4}\right)D\left(F_{ab}\star F\right)\nonumber \\
 & \quad\quad\quad+\left(\frac{\gamma_{2}}{4}+\frac{\gamma_{3}}{2}\right)D\iota_{ba}\left(F\wedge\star F\right).\label{eq: connection field equation}
\end{align}
This equation may be solved algebraically for the Lagrange multiplier
1-forms. The solution reads 
\begin{align}
\lambda_{a} & =-\frac{4}{\mu}Y_{a}+\left(\gamma_{1}+\frac{\gamma_{2}}{2}\right)\left[\iota^{b}D\left(F_{ba}\star F\right)-\frac{1}{4} \iota^{cb}D\left(F_{bc}\star F\right)e_{a}\right]\\
 & \quad\quad+\left(\frac{\gamma_{2}}{2}+\gamma_{3}\right)\left[\iota^{b}D\iota_{ab}\left(F\wedge\star F\right)-\frac{1}{4}\iota^{cb}D\iota_{cb}\left(F\wedge\star F\right)e_{a}\right],\label{lagmult}
\end{align}
where $Y_{a}=Ric_{a}-(1/4)Re_{a}$ are the Schouten 1-forms. Finally,
the co-frame variations produce the Einstein field equations
\begin{align}
0 & =\frac{1}{2}\epsilon_{abc}R^{bc}-\frac{\Lambda}{2}\epsilon_{abc}e^{bc}-\frac{1}{2}\iota_{a}F\wedge\star F+\frac{1}{2}\left(\iota_{a}\star F\right)F+D\lambda_{a}\nonumber \\
 & \quad+\left(\frac{\gamma_{1}}{2}+\frac{\gamma_{2}}{4}\right)\bigg[\frac{1}{2}F^{bc}\left(\iota_{a}\star R_{bc}\right)F-2F_{a}^{\phantom{a}b}Ric_{b}\wedge\star F\nonumber \\
 & \quad\quad\quad\quad\quad\quad\quad\quad\quad+2F_{a}^{\phantom{a}b}\left(\iota^{c}\star F\right)R_{bc}+\frac{1}{2}F^{bc}\left(\iota_{a}\star F\right)R_{bc}\bigg]\nonumber \\
 & \quad\quad+\left(\frac{\gamma_{2}}{4}+\frac{\gamma_{3}}{2}\right)\Big[-\epsilon_{abc}\star\left(F\wedge\star F\right)R^{bc}-2Ric_{a}F\wedge\star F\Big],\label{eq: einstein field equations}
\end{align}
where the Lagrange multiplier 1-forms are given by the expression
(\ref{lagmult}). We note that when the non-minimal coupling constants are taken to be zero, the Einstein field equations (\ref{eq: einstein field equations}) and modified Maxwell's equations (\ref{eq: modified maxwell}) become that of TMG with electrodynamics as expected. Now, we will be looking for exact solutions of these equations. 

\section{$\text{AdS}_{3}$ solution with constant electromagnetic field}

We will be working on constant negative curvature background, that
is $\mathrm{AdS}_{3}$ space-time. We will be working with the left-invariant
1-forms on $\mathrm{AdS}_{3}\cong\mathrm{SO}(2,2)/\mathrm{SO}(2,1)$,
which satisfy 
\begin{equation}
\d e^{a}=-\frac{1}{\rho}\epsilon_{\phantom{a}bc}^{a}e^{b}\wedge e^{c},
\end{equation}
where $\rho=(-\Lambda)^{-1/2}$ is the AdS radius.\footnote{A detailed account of this coordinate system may be found in the Appendix of \cite{DereliYetismisoglu}.} In this coordinate
system Levi-Civita connection 1-forms read 
\begin{equation}
\omega_{\phantom{a}b}^{a}=-\frac{1}{\rho}\epsilon_{\phantom{a}bc}^{a}e^{c}.
\end{equation}
Finally, the curvature 2-forms, Ricci 1-forms and curvature scalar
are given by 
\begin{equation}
R_{\phantom{a}b}^{a}=-\frac{1}{\rho^{2}}e^{a}\wedge e_{b},\quad Ric_{a}=-\frac{2}{\rho^{2}}e_{a},\quad R=-\frac{6}{\rho^{2}},
\end{equation}
respectively. For the electromagnetic field, we take the ``self-dual''
ansatz
\begin{equation}
F=-Ee^{01}+Be^{12},
\end{equation}
with $E=kB$ where $k^{2}=1$. Furthermore we assume that the electric and magnetic fields are constant. These assumptions give, in particular, $F\wedge\star F=0$ which leads to great simplifications. After a tedious computation, 
Einstein field equations (\ref{eq: einstein field equations}) and
modified Maxwell's equation (\ref{eq: modified maxwell}) become algebraic
relations in terms of the AdS radius and coupling constants: 
\begin{align}
2\rho^{2}+14\gamma_{1}+\gamma_{2}-12\gamma_{3}=0,\label{eq: rho constraint}\\
m\rho^{3}+2\rho^{2}+4\gamma_{1}+8\gamma_{2}+12\gamma_{3}=0.\label{eq: m constraint}
\end{align}
These equations effectively carve out a three dimensional subspace
of the parameter space $\{\rho,m,\gamma_{1},\gamma_{2},\gamma_{3}\}$.
To be concrete, we can take the coupling constants $\gamma_{1},\gamma_{2},\gamma_{3}$
for the non-minimal couplings to be arbitrary and they completely
determine the AdS radius and electromagnetic Chern-Simons coupling
constant. Although the standard values of the parameters $\gamma_i$ are calculated in four dimensions, to get a rough idea one may calculate the orders of magnitude of $\rho$ and $m$ using these values. This yields \textcolor{black}{
\begin{align*}
\rho & \sim10^{-2}\lambdabar_{\text{e}}\sim10^{-14}\ \mathrm{m},\\
m & \sim10\cdot m_{\text{e}}\sim10^{-29}\ \mathrm{kg},
\end{align*}
where $\lambdabar_{\text{e}}=\hbar/m_{\text{e}}c$ is the reduced
Compton wavelength of the electron.}

\section{Conclusion}
Starting from the action formulation of topologically massive gravity with electrodynamics extended by the non-minimal coupling terms, we derived variational field equations. We showed that there exists exact constant negative curvature background with constant self-dual electromagnetic fields provided that the
parameters of the model satisfy the constraints \eqref{eq: rho constraint}-\eqref{eq: m constraint}. Although the solution given in this work is a highly restrictive one, the existence of a solution is theoretically stimulating since this shows, in particular, that the model is not inconsistent. Moreover, this simple solution may serve as a starting point for looking for
more general solutions. The solution we have found is similar to the work of Dereli and \c{S}eniko\u{g}lu \cite{DereliSenikoglu}, where the existence of a spherically symmetric, asymptotically de Sitter solution in four dimensions with a constant electromagnetic field 2-form was shown, for a simple $RF^{2}$-type
non-minimal interaction.

\ack{The authors would like to thank Prof. Tekin Dereli, in particular for stimulating discussions on three dimensional gravity models and in general for his everlasting guidance. A similar model, which is a subcase of the model considered here, was suggested by Tekin Dereli to one of us (CY) as an independent study problem. Although the independent study was completed, the results were never published.}

\section*{References}
\providecommand{\newblock}{}


\end{document}